\documentclass[aps,prd,nofootinbib,twocolumn,superscriptaddress,letterpaper,preprintnumbers]{revtex4-1}

\usepackage{amsmath, amssymb,braket, mathtools}
\usepackage[dvipsnames]{xcolor}
\usepackage{color}
\usepackage{float}
\usepackage{graphicx}
\usepackage{subfigure}
\usepackage{natbib}
\usepackage[colorlinks=true
,urlcolor=blue
,anchorcolor=blue
,citecolor=blue
,filecolor=blue
,linkcolor=red
,menucolor=blue
,hyperfootnotes=false,
,linktocpage=true
,pdfproducer=medialab
,pdfa=true
]{hyperref}

\newcommand{\phidot}{\dot{\phi}_0}
\newcommand{\D}{{\rm d}}
\newcommand{\mt}{{\tilde{m}}}

\newcommand{\fnl}{f_{\rm NL}}
\newcommand{\cs}{{\cal{S}}}
\let\vec\mathbf
\newcommand{\es}[2] {\begin{equation} \label{#1} \begin{split} #2 \end{split} \end{equation}}

\newcommand{\dd}{{\rm d}}

\newcommand{\hinf}{H}

\begin{document}
\title{Cosmological Collider Searches beyond the Hubble Scale with {\it Planck} Data}

\author{Soubhik Kumar} 
\affiliation{Institute of Cosmology, Department of Physics and Astronomy, Tufts University, Medford, MA 02155, USA}
\author{Qianshu Lu} 
\affiliation{School of Natural Sciences, Institute for Advanced Study, Princeton, NJ 08540, USA}
\affiliation{Center for Cosmology and Particle Physics, Department of Physics, New York University, New York, NY 10003, USA}
\author{Zhong-Zhi Xianyu}
\affiliation{Department of Physics, Tsinghua University, Beijing 100084, China}
\affiliation{Peng Huanwu Center for Fundamental Theory, Hefei, Anhui 230026, China }
\author{Yisong Zhang}
\affiliation{Department of Physics, Tsinghua University, Beijing 100084, China}
\begin{abstract}
Searches for primordial non-Gaussianity (NG) has the potential to not only reveal the physics of cosmic inflation, but also the structure of fundamental interactions at the highest energies.
The cosmological collider (CC) physics program exemplifies this possibility and demonstrates how searches for {\it oscillatory} NG can lead to mass-spin {\it spectroscopy} of extremely heavy states.
Adopting an effective field theory approach, we find the class of Feynman diagrams that can give the largest NG mediated by a heavy scalar particle with mass $M\sim H$, the inflationary Hubble scale.
We compute the full shape of the NG and perform the first search for this shape using {\it Planck} data, finding no evidence for NG.
This search loses its sensitivity as $M\gg H$ since quantum vacuum fluctuations cannot efficiently produce such heavier particles.
We then focus on a mechanism where a {\it chemical potential} excites {\it on-shell} scalar particles with mass $M\gg H$.
Computing the full shapes, we perform the first CC search for particles parametrically heavier than $H$ using {\it Planck} data. For a range of chemical potential $\omega$ and $M$ satisfying $\omega-M \simeq 3H$, we find a global $1.7\sigma$ evidence for non-zero NG, after taking into account the look-elsewhere effect.
\end{abstract}
\maketitle
\textbf{Introduction and Summary}--- Cosmic inflation is the leading paradigm that can explain the observed large-scale density fluctuations in our Universe~\cite{Baumann:2009ds}.
A central challenge of modern cosmology is to understand the microphysics of this inflationary epoch~\cite{Achucarro:2022qrl}.
Precision searches for correlation functions of cosmological fluctuations are particulary powerful in this regard since those can extract a variety of key properties, including the symmetries governing inflationary dynamics and the (self-)interactions of the inflationary fluctuations~\cite{Chen:2010xka, Wang:2013zva, Green:2022bre}.
Remarkably, the energy scale of inflation can be as high as $10^{16}~{\rm GeV}$~\cite{BICEP:2021xfz}, and such cosmological correlators can reveal the fundamental physics operating at these extremely high scales where some of the deepest ideas ranging from gauge coupling unification to extra spatial dimensions to quantum gravity could reside.

This aspect has been increasingly appreciated in the past decade within the program of cosmological collider (CC) physics~\cite{Chen:2009we, Chen:2009zp, Baumann:2011nk, Chen:2012ge,Pi:2012gf,Gong:2013sma,Arkani-Hamed:2015bza}.
States that are heavier than the inflationary Hubble scale $\hinf$ can be produced during inflation and leave distinct {\it oscillatory} signatures in various soft limits of $n$-point correlators $(n\geq 3)$.
It has been demonstrated, together with several explicit mechanisms and model realizations~\cite{Baumann:2011su, Assassi:2013gxa, Craig:2014rta,  Dimastrogiovanni:2015pla, Lee:2016vti, Meerburg:2016zdz, Chen:2016uwp, Chen:2016nrs, Chen:2016hrz, An:2017hlx, Chen:2017ryl, Kumar:2017ecc, Baumann:2017jvh,  Chen:2018xck, Kumar:2018jxz, Wu:2018lmx, Dimastrogiovanni:2018uqy, Lu:2019tjj, Hook:2019zxa, Hook:2019vcn, Kumar:2019ebj, Wang:2019gbi, Li:2019ves, Alexander:2019vtb, Bodas:2020yho, Wang:2020ioa, Lu:2021gso, Lu:2021wxu, Dimastrogiovanni:2021cif, Cui:2021iie, Tong:2022cdz, Qin:2022lva, Reece:2022soh, Chen:2022vzh, Maru:2022bhr, Chen:2023txq,   Craig:2024qgy, Quintin:2024boj, Bodas:2024hih, Hubisz:2024xnj, Chakraborty:2025myb, deRham:2025mjh, Bodas:2025vpb,  Chakraborty:2025mhh, Aoki:2025uff, Kumar:2025anx, Jiang:2025mlm}, that a measurement of such oscillations can reveal key information about those heavy states, including their masses and spins.

Meanwhile, cosmic microwave background (CMB) and large-scale structure (LSS) observations are steadily gathering new data~\cite{Planck:2019kim, Jung:2025nss, Chaussidon:2024qni, SPHEREx:2014bgr,SimonsObservatory:2018koc, EUCLID:2024yrr}, using which CC predictions of many realistic mechanisms can now be tested. 
For this purpose, it is important to confront precision data with the {\it full} shape of the $n$-point correlator, including both the oscillatory \emph{signal} and the relatively featureless \emph{background}, since both are generally present in realistic models. 
However, this has been challenging due to several technical difficulties ranging from theoretical computation to data analysis, and progress has started only very recently. 
On the theory front, new analytical methods, such as `cosmological bootstrap'~\cite{Arkani-Hamed:2018kmz,Baumann:2019oyu,Pimentel:2022fsc,Jazayeri:2022kjy,Qin:2022fbv,Qin:2023ejc,Aoki:2023wdc,Aoki:2024uyi,Chen:2024glu,Liu:2024str}, and numerical methods, such as efficient evaluation of nested integrals~\cite{Wang:2021qez,Werth:2023pfl,Pinol:2023oux,Werth:2024aui},  have been developed to generate full shapes of three-point and four-point correlators in several scenarios.
New analysis pipelines~\cite{Sohn:2023fte, Philcox:2025bvj,Philcox:2025lrr, Zhang:2025nyb} are also being developed to search for realistic correlators that are not factorizable into functions of individual momenta. 

\begin{figure}
    \centering
    \includegraphics[width=0.95\linewidth]{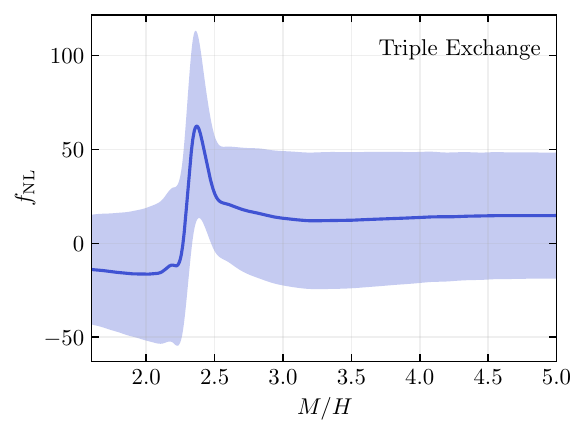}
    \includegraphics[width=0.95\linewidth]{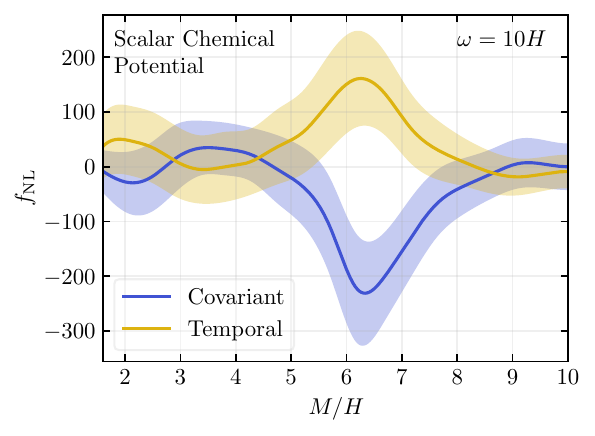}
    \caption{Results of the search performed in this work with {\it Planck} data. {\it Top:} Estimated strength of the bispectrum ($\fnl$) mediated by a heavy scalar with mass $M$ via the `triple-exchange' diagram. This search is consistent with $\fnl=0$ at 95\% CL. {\it Bottom:} Estimated $\fnl$ in the `scalar chemical potential'~\cite{Bodas:2020yho} scenario for two different shape functions originating from temporal and covariant derivative couplings. For the latter and for a chemical potential $\omega=10\hinf$ and $M=6.7H$, we find $\fnl =  -203 \pm 82$ at 68 \% CL, with a $2.5\sigma$ local, corresponding to $1.7 \sigma$ global significance for $\fnl\neq 0$. Similar significances are obtained for other combinations satisfying $\omega-M \simeq 3H$. The error bands represent 68\% CL. See the main text for further details.}
    \label{fig:fnl_bound}
\end{figure}
Indeed, several recent works have searched for CC signals in cosmological data~\cite{Green:2023uyz, Cabass:2024wob, Goldstein:2024bky, Sohn:2024xzd, Suman:2025vuf, Suman:2025tpv, Philcox:2025wts, Bao:2025onc, Anbajagane:2025uro, Philcox:2025bbo, Green:2026yev}, using certain types of couplings between the inflaton and heavy states, and primarily focusing on `single-exchange' processes where only one heavy particle is exchanged between the inflaton quanta. 
However, effective field theory (EFT) considerations allow for more general couplings that naturally lead to other processes with parametrically larger three- and four-point functions.
In fact, some of these couplings were already present in the original quasi-single-inflation (QSFI) model~\cite{Chen:2009we, Chen:2009zp}.

In this \emph{Letter}, we present the \emph{full-shape} searches for realistic CC scenarios, involving generic tree-level scalar particle exchanges, by looking for three-point correlators (bispectrum) in the {\it Planck} 2018 data. 
Specifically, we search for the first time the bispectrum mediated by the so-called `triple-exchange' diagram that contains three massive scalar propagators and often gives the largest bispectrum, especially within {\it Planck} sensitivity, compared to other contributions previously searched for.
To perform this search, we use the {\tt CMB-BEST} package~\cite{Sohn:2023fte}, which automates the computation of numerically expensive steps in the data analysis, along with our evaluation of the full shape function. We also use the correlation matrix from \texttt{CMB-BEST} for each model across the mass range we searched for to correct for the look-elsewhere effect following~\cite{Sohn:2024xzd}.
We do not find any evidence for bispectrum in this model-independent search (Fig.~\ref{fig:fnl_bound}, top) and interpret our results to place the first constraint on the QSFI model.

\begin{figure*}
    \centering
    \includegraphics[width=0.9\linewidth]{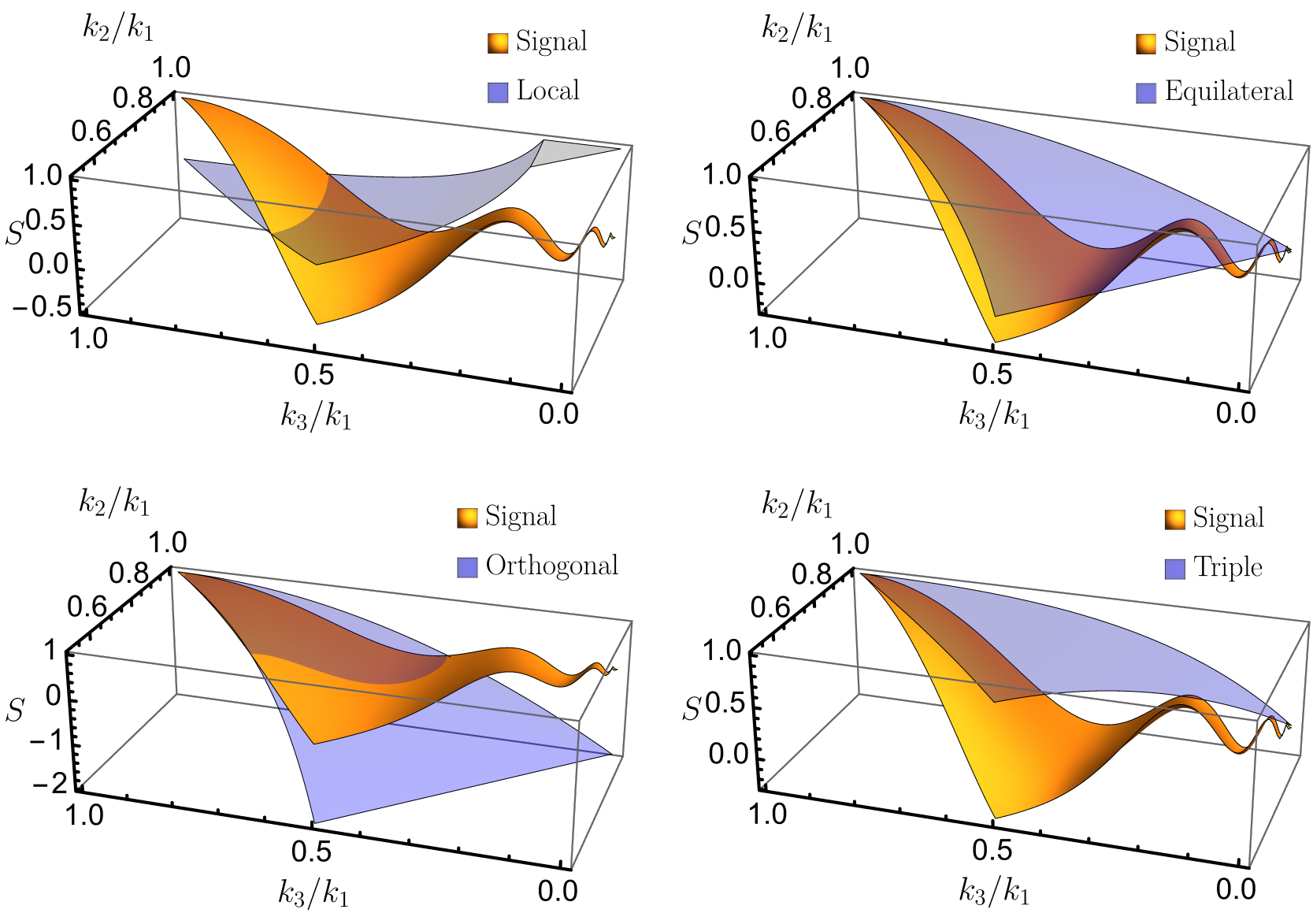}
    \caption{Comparison of $\cs^{\rm cov}(\omega=10H, M=6.7H)$ (`Signal') with the local (top left), equilateral (top right), orthogonal (bottom left), and $\cs^{\rm triple}(M=6.7H)$ (bottom right) shapes. The cosine correlation~\cite{Sohn:2024xzd} between the `Signal' and the four shapes calculated by \texttt{CMB-BEST} are 5\%, 54\%, 22\%, and 20\% respectively.}
    \label{fig:shape_comp}
\end{figure*}

In the above scenario, as well as in all the previous searches in the literature, the distinctive oscillatory signature gets exponentially suppressed as the heavy state mass $M$ increases beyond $\hinf$.
This limits the reach of the CC searches with current data to $M \lesssim 2\hinf$.
Several mechanisms have been proposed that produce states parametrically heavier than $\hinf$ and extend the reach of the CC~\cite{Chen:2018xck, Hook:2019zxa, Bodas:2020yho, Chen:2022vzh, Wang:2020ioa}.
We focus on one such mechanism, the `scalar chemical potential'~\cite{Bodas:2020yho}.
We compute the associated {\it full-shape} bispectrum for the first time and search for oscillatory and unsuppressed signals from particles as heavy as $M\simeq 10\hinf$ using {\it Planck} data (Fig.~\ref{fig:fnl_bound}, bottom).
For a range of chemical potential $\omega$ and mass values satisfying $\omega-M \approx 3H$, we get a $2.5\sigma$ local, corresponding to $1.7\sigma$ global significance for $\fnl\neq 0$ for the `covariant shape', discussed below. In particular, for $\omega=10\hinf$ and $M=6.7H$, we get a bispectrum strength $\fnl = -203 \pm 82$ at 68 \% CL (defined in eq.~\eqref{eq:fnl}), with a $2.5\sigma$ local significance for $\fnl\neq 0$. Importantly, the full shape for this parameter point does not resemble any standard shapes used for previous bispectrum searches (see Fig.~\ref{fig:shape_comp}).
In particular, the sensitivity is driven by the first peak near the equilateral region~\cite{paper_2}, where a simplified signal-only template is inapplicable.

\textbf{Heavy Scalars in the Cosmological Collider}--- The bispectrum of curvature fluctuation $\zeta$ is conventionally parameterized by a dimensionless shape function $\cs$:
\es{eq:fnl}{
\langle \zeta(\Vec{k}_1) \zeta(\Vec{k}_2) \zeta(\Vec{k}_3)\rangle' \equiv {18 \over 5} \fnl (2\pi^2 A_s)^2 {\cs(k_1, k_2, k_3) \over k_1^2 k_2^2 k_3^2},
}
where $A_s \equiv (k^3/(2\pi^2)) (k_*/k)^{n_s-1}\langle\zeta(\Vec{k})\zeta(-\Vec{k})\rangle'$ is the dimensionless power spectrum and $k=|\vec{k}|$.
Here, we have used the notation $\langle \cdots\rangle = (2\pi)^3 \delta^3(\sum \Vec{k}_i)\langle \cdots\rangle'$.
The kinematic dependence of the CC signal is fully encoded in $\cs$, while its strength is encoded in $\fnl$.
A specific normalization of $\cs$ is required to unambiguously quantify $\fnl$ and will be discussed below. 
While a variety of models and mechanisms have been explored (see~\cite{Wang:2019gbi} for a summary) to understand CC signatures, here we adopt an EFT-based framework by focusing on slow-roll inflation.
This allows our searches to be agnostic to the detailed inflationary potential while being easily interpretable in the context of full models, including the QSFI model.
In our set up, theoretical control requires the EFT cut off $\Lambda > \phidot^{1/2}$~\cite{Creminelli:2003iq}, the slow-roll velocity of the inflaton.
This means a non-trivial speed of sound $c_s$, originating from a dimension-8 operator $(\partial\phi)^4/\Lambda^4$, with $c_s^2 =1/(1+4\phidot^2/\Lambda^4)$, is close to unity.
However, it is straightforward to extend our analysis to the even more agnostic framework of Goldstone EFT of inflation~\cite{Cheung:2007st}, which includes $c_s\ll 1$.

In our EFT description, we extend single-field slow-roll inflation by including an additional heavy scalar field $\sigma$ which mediates CC signals at the tree level.
Respecting the shift symmetry of the inflaton field $\phi$, the leading terms of an EFT expansion is given by,
\es{eq:EFT}{
{\cal L} \supset  -\frac12 M^2 \sigma^2 - \frac16 \kappa \sigma^3 + {(\partial \phi)^2\sigma\over \Lambda_1} + {(\partial \phi)^2\sigma^2\over \Lambda_2^2}.
}
Here $(\partial \phi)^2 = g^{\mu\nu}\partial_\mu\phi\partial_\nu\phi$ with the spacetime metric $\D s^2 = g_{\mu\nu}\D x^\mu \D x^\nu =  -\D t^2 + a(t)^2 \D \vec{x}^2$.
The bare mass $M$ and the coupling coefficient $\kappa$ along with EFT cutoffs $\Lambda_1 \sim \Lambda_2$ may be related in a specific model but we leave them independent. 
During inflation, $\phi$ acquires a slow-roll background $\phi_0(t)=\phi_*+\dot\phi_0t+\cdots$ where $\dot\phi_0=H^2/(2\pi\sqrt{A_s})\approx (59H)^2$ from the best-fit value of $A_s$~\cite{Planck:2018vyg, AtacamaCosmologyTelescope:2025blo}.
Consequently, in terms of $\sigma$ and inflaton fluctuations $\varphi\equiv \phi-\phi_0(t)$, the modification to the free Lagrangian is given by
\es{eq:expanded}{
{\cal L} \supset & -{\kappa\over 6}\sigma^3-\frac{2\dot\phi_0}{\Lambda_1}\dot{\varphi}\sigma 
+\frac{(\partial\varphi)^2}{\Lambda_1}\sigma-\frac{2\dot\phi_0}{\Lambda_2^2}\dot{\varphi}\sigma^2,
}
where the dots denote derivatives with respect to $t$.
While we have not explicitly shown it in eq.~\eqref{eq:expanded}, the dimension-six operator in eq.~\eqref{eq:EFT} gives a tree-level correction to $\sigma$ mass $\Delta M^2 = 2\phidot^2/\Lambda_2^2$.
 
These EFT couplings naturally allow three possible tree-level graphs: single-, double- and triple-exchange diagrams, depending on the number of $\sigma$ propagators: 
\es{}{
\includegraphics[width=0.47\textwidth]{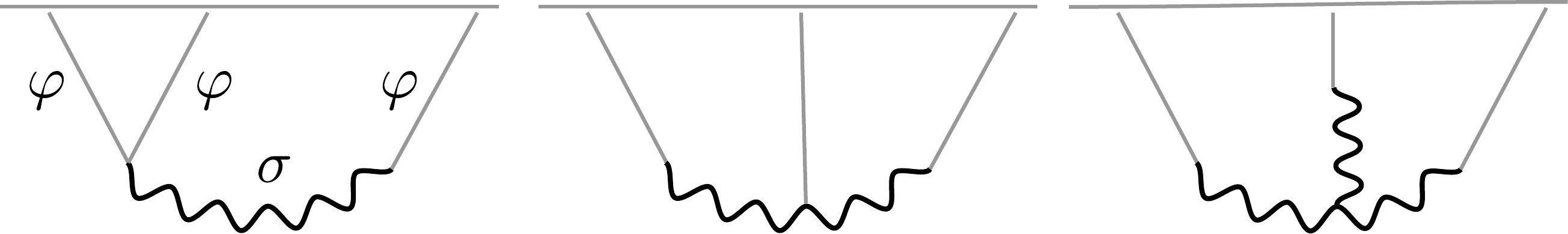}\nonumber
} 
Previous observational searches have primarily focused on the single-exchange, and for limited benchmarks, the double-exchange diagram~\cite{Philcox:2025bbo}. 
In this {\it Letter}, we perform the first search for bispectrum from the triple-exchange diagram, leaving the full calculation and detailed searches including single-  and double-exchange diagrams for a companion work~\cite{paper_2}.
 
The triple-exchange diagram can naturally lead to a parametrically enhanced bispectrum relative to single and double-exchange diagrams for two reasons. 
First, the quadratic mixing between $\varphi$ and $\sigma$ in eq.~\eqref{eq:expanded} corrects the power spectrum, $\Delta P_\zeta/P_\zeta =  \mathcal{P}(M) \phidot^2/(\Lambda_1^2 H^2)$, where $\mathcal{P}(M)\sim 0.03 - 1.3$ for $M \in [1.5, 2.5]$ and asymptotes to $H^2/M^2$ for larger $M$~\cite{Chen:2012ge, Chen:2017ryl}. We require $\Delta P_\zeta/P_\zeta<1$ (this can be relaxed, as discussed below). 
Second, if $\Delta M^2 > M^2 \gtrsim H^2$, the effective mass of $\sigma$ would be too large to give observable oscillatory bispectrum.
The absence of tree-level tuning between $M^2$ and $\Delta M^2$  requires ${\Delta M^2 / M^2} \sim {\phidot^2 / (M^2 \Lambda_2^2)} < 1$.
Imposing these inequalities, we estimate the sizes of the oscillatory signals from single-, double-, and triple-exchange processes \cite{Xianyu:2025lbk}:
\begin{equation}
\begin{split}
& f^{\rm osc}_{\rm NL, si} \sim \frac{(\Delta P_\zeta / P_\zeta)}{\mathcal{P}(M)}  (M/H)^{3\over 2}~{\rm Exp}<1,\\
&f^{\rm osc}_{\rm NL, do} \sim \frac{(\Delta P_\zeta / P_\zeta)}{\mathcal{P}(M)} (M/H)^{-{1\over 2}}(\Delta M/H)^2 ~{\rm Exp}<1,\\
& f^{\rm osc}_{\rm NL, tr} \sim \frac{(\Delta P_\zeta / P_\zeta)^{3 \over 2}}{{(\mathcal{P}(M))^{3\over 2}}} (M/H)^{- {5\over 2}} {\kappa \phidot \over H^3} ~{\rm Exp}
<3 \times 10^3,\label{eq:est_xexch}
\end{split}\raisetag{1\baselineskip}
\end{equation}
where ${\rm Exp}=\exp(-\pi M/H)$, and in the last relation we have used $M\sim \mathcal{O}(H)$, $\kappa\sim 4\sqrt{\pi}M$ at the partial wave unitarity limit, and $\dot\phi_0^{1/2}\approx 59H$.
This shows that the oscillatory shape from the triple-exchange process can be parametrically larger than the other two, and would be readily probed by {\it Planck} data which is sensitive to $\fnl \gtrsim {\cal O}(10-100)$.
Relaxing $\Delta P_\zeta/P_\zeta<1$ requirement does not change this conclusion. The $\Delta P_\zeta/P_\zeta$ in the first two of eq.~\eqref{eq:est_xexch} is simply replaced by $P_\zeta/P^{\rm new}_\zeta$, where $P_\zeta \sim H^4/\phidot^2$ is the power spectrum without mixing, and $P^{\rm new}_\zeta \sim (\phidot/(\Lambda_1 M))^2 \times H^4/\phidot^2 \sim H^4/(\Lambda_1 M)^2$ is that with mixing.
In the last one of eq.~\eqref{eq:est_xexch}, $(\Delta P_\zeta/P_\zeta)^{3/2}$ is replaced by $(P_\zeta/P_\zeta^{\rm new})^{1/2}$.
Since $P_\zeta/P_\zeta^{\rm new} \sim M^2\Lambda_1^2/\phidot^2 < 1$, the triple-exchange diagram is again expected to be most relevant in the context of {\it Planck} data.
Therefore, it is crucial to include the triple-exchange diagram when performing the CC searches for massive scalars.
Before discussing the results of this search, we describe the theoretical framework of the second class of searches performed in this work.

\textbf{Cosmological Collider with a Chemical Potential}--- A common feature of the oscillatory signals as shown above is a `Boltzmann factor' $\exp(-\pi M/H)$, due to the fact that quantum fluctuations with size $\sim H$ cannot efficiently produce on-shell particles with $M\gg H$.
This suppresses the signal for $M$ even mildly larger than $H$ and thus severely limits the energy reach of CC physics. 
The suppression can be relieved utilizing the kinetic energy of the inflaton $\phidot^{1/2}\approx 59H$ which can excite particles with masses $H \ll M < \phidot^{1/2}$ and potentially extend the reach of CC physics by almost two orders of magnitude. While this mechanism can work for scalars~\cite{Chen:2022vzh, Bodas:2020yho}, fermions~\cite{Chen:2018xck, Hook:2019zxa}, and gauge bosons~\cite{Wang:2020ioa, Bodas:2024hih}, in this {\it Letter} we focus on the case of a heavy scalar, in particular the `scalar chemical potential' mechanism~\cite{Bodas:2020yho}.

The associated Lagrangian for a complex scalar $\chi$ is,
\es{eq:scp}{
{\cal L} \supset & -|\partial\chi|^2-M^2|\chi|^2+ \mt^3(\chi+\chi^\dagger)\\
&-{i\partial_\mu\phi \over \Lambda}J^\mu-{c(\partial\phi)^2\over \Lambda^2}|\chi|^2,
}
where $J^\mu = \chi\partial^\mu\chi^\dagger-\chi^\dagger\partial^\mu\chi$ is the $U(1)$ current.
The tadpole term softly breaks the $U(1)$ symmetry and leads to a homogeneous classical background for the $\chi$ field $\chi_0 = -\mt^3/(c\omega^2 - M^2+i \Box \phi_0/\Lambda)$ with $\omega \equiv \phidot/\Lambda$ denoting the chemical potential.
Expanding $\phi$ as before and $\chi$ around this VEV $\chi = \chi_0 + \delta\chi$, we obtain the following interaction terms~\cite{Bodas:2025vpb},
\es{eq:scp_expanded}{
\mathcal{L}_{\delta\chi}\supset \frac{c}{\Lambda^2}\left(2\dot{\phi}_0\dot{\varphi}-\left(\partial\varphi\right)^2\right)\left(\chi_0^{\dagger}\delta\chi e^{-i\phi/\Lambda}+{\rm h.c.}\right).
}
Along with a quadratic mixing, the above terms give two types of cubic vertices upon expanding $e^{-i\phi/\Lambda}$:
\es{eq:scp_cubic}{
{\rm temporal~derivative:} -\frac{2i c \omega \chi_0^{\dagger}}{\Lambda^2}\dot{\varphi}\varphi\delta\chi e^{-i \omega t}+{\rm h.c.},\\
{\rm covariant~derivative:}-\frac{c \chi_0^{\dagger}}{\Lambda^2}(\partial\varphi)^2\delta\chi e^{-i \omega t}+{\rm h.c.}.
}
Both types of cubic vertices are present in the full model.
However, we focus on each type separately to illustrate the differences in the associated shape functions ${\cal S}$. For both cases, the rapidly oscillating $\exp(-i\omega t)$ coupling removes the exponential suppression for $\omega > M$ and extends the reach of the CC physics. 
Furthermore, the particular implementation in eq.~\eqref{eq:scp} does not allow for a cubic interaction of $\chi$, and hence in this scenario, the `single-exchange' diagram is the only tree-level contribution.

Having described the theoretical framework of bispectrum mediated by (1) {\it neutral} heavy scalar, where the `triple-exchange' diagram dominates, and (2) a {\it charged} heavy scalar in the presence of a chemical potential, where only `single-exchange' diagram is relevant, we now focus on the full shape functions $\cs$ and then the results from our {\it Planck} searches. 

\textbf{Full Shape Functions}---
As stressed before, it is important to include the full shape function $\cs$ for an accurate search of CC signatures. 
These shape functions are computed from a set Feynman rules in the Schwinger-Keldysh (`in-in') formalism. 
In particular, an $n$-point correlator $\langle \varphi(\vec{k}_1)\cdots \varphi(\vec{k}_n)\rangle \equiv \langle Q\rangle $ at some time $t_0$ towards the end of inflation is computed via 
\es{eq:inin}{
    \langle Q\rangle = \langle 0  | U^\dagger Q_I(t_0) U |0 \rangle,
}
where $U = T\exp(-i\int_{-\infty(1-i\epsilon)}^{t_0} \D t \, \mathcal{H}_I^{\rm int}(t))$ denotes the time evolution operator in terms of the interacting part of the interaction picture Hamiltonian, $\mathcal{H}^{\rm int}_I$, and $|0 \rangle$ is the free vacuum. Determining the in-in expectation value often requires performing multiple nested time integrals over a product of special functions; in our case, Hankel functions.
For example, evaluating the triple-exchange diagram requires computing four-layered integrals due to the presence of four vertices.
This has prohibited an analytic evaluation of this diagram. 

We compute this diagram via direct numerical integration over time with the `coupled mode function' method~\cite{An:2017hlx}, where the quadratic mixing term is included in the equations of motion (EOM) satisfied by $\varphi$ and $\sigma$, and those {\it coupled} EOM are solved numerically.
The quadratic mixing vertex is thus accounted for non-perturbatively, and a mixed propagator $\langle\sigma_\vec{k}(t)\varphi_{-\vec{k}}(t_0)\rangle$ is obtained. The triple-exchange diagram is then reduced to a contact cubic interaction involving three mixed propagators. 
The in-in evaluation then requires a single, non-nested integration which can be done relatively easily to obtain a numerical shape function $\cs^{\rm triple}$.
The explicit EOMs and the form of the mixed propagator can be found in the End Matter.

For the scalar chemical potential scenario, only the single-exchange diagram contributes, as discussed above.
Consequently, we adopt the analytical formalism of~\cite{Qin:2023ejc} and appropriately generalize to the vertices given in eq.~\eqref{eq:scp_expanded}.
We separately compute the contributions from the temporal derivative and the covariant derivate couplings (eq.~\eqref{eq:scp_cubic}) to get $\cs^{\rm temp}$ and $\cs^{\rm cov}$.
The explicit form of the three-point function $\langle \varphi(\vec{k}_1)\varphi(\vec{k}_2)\varphi(\vec{k}_3)\rangle$ determining these shape functions can be found in the End Matter.

To perform an accurate search, we need sufficiently precise shape functions.
An evaluation on a fine 3D grid in the $(k_1, k_2, k_3)$ is computationally expensive.
However, we exploit scale invariance to first evaluate all the three shape functions $\cs^{\rm triple}$, $\cs^{\rm temp}$, and $\cs^{\rm cov}$, with a uniform $\log_{10}$ grid of 120 points in each direction in the 2D $(k_2/k_1, k_3/k_1)$ plane. These are then appropriately extended, via cyclic symmetry, to the entire $(k_1, k_2, k_3)$ cube corresponding to a total of 216,000 lattice points. We focus on the \texttt{CMB-BEST} dynamical range $10^{-3}\leq k_1/k_{\rm max}, k_2/k_{\rm max}, k_3/k_{\rm max} \leq 1, k_{\rm max} = 0.209/{\rm Mpc}$, and repeat these steps for each model parameter.
In particular, for $\cs^{\rm triple}$ we consider 50 equally spaced values of the scalar mass in the range $1/10\leq \tilde{\nu}\leq 5$ where $\tilde{\nu} = (M^2/H^2-9/4)^{1/2}$.
For each of $\cs^{\rm temp}$ and $\cs^{\rm cov}$, we consider $1\leq \omega/H\leq 10$, in steps of 1, along with $1/2\leq \tilde{\nu}\leq 10$ in steps of 1/2.
\begin{figure}
    \centering
    \includegraphics[width=0.95\linewidth]{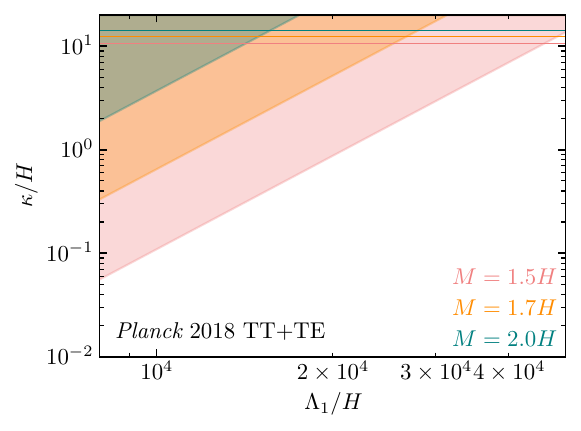}
    \caption{Constraints on the QSFI model from the {\it Planck} search for bispectrum mediated by the triple-exchange diagrams. The bispectra search probes heavy scalars for $1.5H\leq M \leq 2H$, while for larger $M$ the bispectrum constraint weakens, and are not shown. The partial wave unitarity bounds $\kappa \leq 4\sqrt{\pi} M$ are violated above the color-coded horizontal lines.}
    \label{fig:triple}
\end{figure}

\textbf{Search with {\it Planck} Data}---
We use the {\tt CMB-BEST} package~\cite{Sohn:2023fte} to perform the {\it Planck} search.
Given an input shape function $\cs_{\rm CMB} \equiv \fnl \cs$, {\tt CMB-BEST} can estimate the bispectrum strength $\fnl$ with its uncertainties. It is common to set $\cs(k_*,k_*,k_*)=1$ for some pivot scale $k_*$. However, we are interested in bispectra that oscillate with the scalar mass, and hence the shape function can vary rapidly in the above equilateral configuration, leading to artificial fluctuations in the estimated $\fnl$.
As such, we normalize $\cs$ such that its maximal value in the entire physical $k_1,k_2,k_3$ volume is unity.
Thus, our input shape function is given by $\cs_{\rm CMB}\equiv \fnl^i \left(k_1^2 k_2^2 k_3^2 B^i(k_1,k_2,k_3)\right)/{\rm Max}\left\{|k_1^2 k_2^2 k_3^2 B^i(k_1,k_2,k_3)|\right\}$ where $B^i(k_1,k_2,k_3) = \langle \varphi(\vec{k}_1)\varphi(\vec{k}_2)\varphi(\vec{k}_3)\rangle'_i$ for $i= {\rm triple}$, temp, cov.

The results for $\fnl^{\rm triple}$ as a function of $M$ is shown in Fig.~\ref{fig:fnl_bound} (top). 
We do not find any evidence of non-zero bispectrum at 95\% CL.
Notably, for $M\approx 2.4H$, $\fnl = 50\pm 40$ at a local $1.25\sigma$.
For this scalar mass, the smooth background component of $\cs^{\rm triple}$ becomes much smaller than the oscillatory signal. This vanishing of the background component at $M\approx 2.4H$ has been shown analytically to leading order in the squeezed limit in~\cite{Xianyu:2025lbk}. Since $\fnl^{\rm triple}$ depends on three parameters, $\kappa, \Lambda_1$, and $M$ (eq.~\eqref{eq:expanded}), we can interpret our search result on the $\{\kappa, \Lambda_1\}$ parameter space for fixed values of $M$ (Fig.~\ref{fig:triple}).
For each $M$, the color-shaded regions are ruled out by our bispectrum search at 68\% CL. Here, we have included the leading correction to the power spectrum from the quadratic mixing term in eq.~\eqref{eq:expanded} in the normalization $A_s$.
For $M>2H$, bispectrum constraints become weaker and are not shown.
For $\Lambda_1 < 8\times 10^3 H$, $P_\zeta^{\rm new} > P_\zeta$, indicating even higher order corrections can be important, and we do not consider such values of $\Lambda_1$.
As is also seen from Fig.~\ref{fig:fnl_bound} (top), the constraints on $\fnl$ becomes independent of $M$ for $M>3H$ since for these values of $M$, the oscillatory signal is too small and the dominant contribution to the bispectrum comes from the analytic background piece.
The $\fnl$ contribution of this piece is independent of $M$ with our normalization convention.

The results for $\fnl^{\rm temp}$ and $\fnl^{\rm cov}$ for $\omega=10 H$ are shown in Fig.~\ref{fig:fnl_bound} (bottom).
While $\fnl^{\rm temp}$ is consistent with zero at 95\% CL, we find $\fnl^{\rm cov}=-203 \pm 82$ for $M\approx 6.7H$, with a 2.5$\sigma$ local, corresponding to 1.7$\sigma$ global evidence for a non-zero $\fnl$.
The associated shape function $\cs^{\rm cov}$ is shown in Fig.~\ref{fig:shape_comp} which illustrates the difference of $\cs^{\rm cov}$ from the local, equilateral, and orthogonal shapes that have been extensively searched for.
We also make a comparison with $\cs^{\rm triple}$ for $M\approx 6.7H$ to illustrate the non-trivial difference from the absence of the exponential suppression due to the chemical potential.
These comparisons illustrate that the distinctive oscillations in ${\cal S}^{\rm cov}$ already set in at mildly equilateral regimes $k_3/k_1 \sim 0.4, k_2 \sim k_1$, and these are not captured by any of the other four shapes.
Thus, in these scenarios full computations of the shape functions and dedicated searches are essential.

\textbf{Model Interpretation}--- We now interpret the results of our chemical potential search using the full model (eq.~\eqref{eq:scp_expanded}), in which both the temporal and the covariant derivative couplings are present. 
The temporal derivative contribution to $\fnl$ dominates over that from the covariant derivative, due to the presence of the additional factor of $\omega$ in the cubic vertex (eq.~\eqref{eq:scp_cubic}).
Thus, we focus on the observational results for the `temporal' shape from the bottom panel of Fig.~\ref{fig:fnl_bound} to understand the implications for the full model.
The associated $\fnl$ is given by
\begin{equation}
    \fnl^{\text{temp}} \sim \frac{\Delta P_{\zeta}}{P_{\zeta}} \frac{\omega {\cal N}}{\mathcal{F}(\omega, M)}\lesssim \frac{\omega {\cal N}}{\mathcal{F}(\omega, M)},\label{eq:CP_fnl_size}
\end{equation}
where ${\cal N}={\rm Max}\left\{|k_1^2 k_2^2 k_3^2 B^{\rm temp}(k_1,k_2,k_3)|\right\}$ is a normalization factor that ensures the maximum of $\cs^{\rm temp}$ is unity.
${\cal N}$ varies from $\sim 1$ at $M\sim H$ to $\sim 0.01$ at $M\sim 10H$ (for $\omega = 10$).
We have also used that the quadratic mixing from~\eqref{eq:scp_expanded} induces a correction to the power spectrum $\Delta P_{\zeta}/P_{\zeta} \sim \mathcal{F}(\omega, M)(2 c \omega/\Lambda)^2 |\chi_0|/(2H^2)$, where $\mathcal{F}(\omega = 10, M)\sim 0.5$ in the full range of $M$. 
We see that demanding $\Delta P_{\zeta}<P_{\zeta}$ implies $|\fnl^{\text{temp}}|\lesssim 20$ to $0.2$ for $M\sim H$ to $10H$ and $\omega = 10$.

Now, as shown in the bottom panel of Fig.~\ref{fig:fnl_bound} (temporal shape), the search for the chemical potential model with $\omega = 10H$ has yielded constraints consistent with null within $1\sigma$ for $M/H\in [1.5, 5.2]$ and $M/H\in [7, 10]$.
For these mass ranges, the observational bound on $|\fnl^{\rm temp}|$ is weaker than the above power spectrum constraint, and ranges from $\sim 60$ to $30$ for $M\sim H$ to $10H$.
This implies demanding $\Delta P_{\zeta}<P_{\zeta}$ provides stronger constraints than the bispectrum search.
In the mass range $M/H\in [5.2, 7]$, we have a mild local significance for a non-zero $\fnl$, peaking at $M/H = 6.7$ at which $\fnl^{\rm temp}=160\pm 88$, with $\fnl\neq 0$ at a global 1.0$\sigma$ (local $ 1.8 \sigma$). 
However, even if this persists with further analysis, the power spectrum bound $|\fnl^{\rm temp}|\lesssim 0.2-20$, discussed before, indicates that the significance cannot be consistently attributed to the chemical potential model with the specific parameter values considered here.

Nonetheless, in other regions of the parameter space the power spectrum bound can be evaded.
As discussed around Fig.~\ref{fig:fnl_bound}, the significance persists across a range of $\omega$ and $M$ values, as long as $\omega - M \approx 3H$, while eq.~\eqref{eq:CP_fnl_size} shows that the expected size of and the upper bound on $\fnl$ increases with $\omega$. 
For example, if we observe a similar significance at $\omega \approx 50 H$ and $M \approx 47 H$, the upper bound would be relaxed by a factor of $\sim 5$, compared with $\omega=10$, and the observed $\fnl^{\rm temp} \sim 100$ could be within this bound.
The scalar chemical potential model would then provide an explanation for the observed significance.
Since our goal here is to present a first search for on-shell effects of $M\gg H$ particles, we leave a detailed scan of the entire parameter space, including $\omega, M \sim 50 H$, for future work.

\textbf{Discussion}---
We stress that the above bound stemming from $\Delta P_{\zeta} < P_{\zeta}$ is not a fundamental one, and we can describe the regime of $\Delta P_{\zeta} > P_{\zeta}$ as long as we `resum' multiple insertions of the $\varphi$-$\chi$ mixing to obtain the full power spectrum. 
Furthermore, eq.~\eqref{eq:CP_fnl_size} again shows the relatively small $\fnl$ from single-exchange diagrams, as already seen in eq.~\eqref{eq:est_xexch}; the existence of a chemical potential counters the $\exp(-\pi M/H)$ suppression, but does not remove the relative $\dot{\phi}_0/H^2$ suppression between single- and triple-exchange diagrams.
However, the model~\eqref{eq:scp} can be simply extended to have triple-exchange diagrams as well.
We leave detailed treatments of these scenarios for future work.

To summarize, we have performed the first search for {\it on-shell} effects of super-$\hinf$ particles using {\it Planck} data. 
Our methodology can be used to perform searches for other CC mechanisms and also with LSS data.
Such searches, especially with upcoming data from SPHEREx~\cite{SPHEREx:2014bgr} and Spec-S5~\cite{Spec-S5:2025uom}, would improve the detection prospects for the searches discussed above.
It would be especially interesting to see if the local $\approx 2.5\sigma$, corresponding to global $\approx 1.7\sigma$, evidence seen in Fig.~\ref{fig:fnl_bound} is also present in LSS data.
Overall, theoretical advances and powerful data analysis techniques have already started to and will further enable us to use cosmological datasets to probe fundamental physics at the highest energies.

\textit{Acknowledgements}---
We thank Arushi Bodas, Xingang Chen, Oliver Philcox, Wuhyun Sohn, Raman Sundrum, and Dong-Gang Wang for useful discussions and comments on a previous draft.
QL is supported in part by the NSF grants PHY-2210498 and PHY-2514611 and by the Simons Foundation. ZX is supported by NSFC under Grants No.\ 12275146 and No.\ 12247103, the National Key R\&D Program of China (2021YFC2203100), and the Dushi Program of Tsinghua University.
This material is based upon work supported by the NSF under grant number 2018149. 
The authors acknowledge the Tufts University High Performance Compute Cluster (https://it.tufts.edu/high-performance-computing) which was utilized for the research reported here.

\newpage

\onecolumngrid
\begin{center}
    \textbf{End Matter}
\end{center}
\twocolumngrid

{\bf Coupled Mode Function Method}---
A direct application of the `in-in' master formula~\eqref{eq:inin} to evaluate the triple-exchane diagram is challenging, both analytically and numerically, since it features four vertices.
Thus, to evaluate it one needs to perform four-layered nested integral involving special functions.
We address this issue by including the quadratic mixing vertex in eq.~\eqref{eq:expanded} in the EOMs of $\varphi$ and $\sigma$.
This way, corrected mode functions for $\varphi$ and $\sigma$, that are non-perturbative in the quadratic mixing, can be derived.
Correspondingly, the triple exchange diagram reduces to a contact diagram with only one vertex, but instead of separate $\varphi$ and $\sigma$ propagators, there are three `mixed' propagators (double solid lines).
\es{}{
\includegraphics[width=0.47\textwidth]{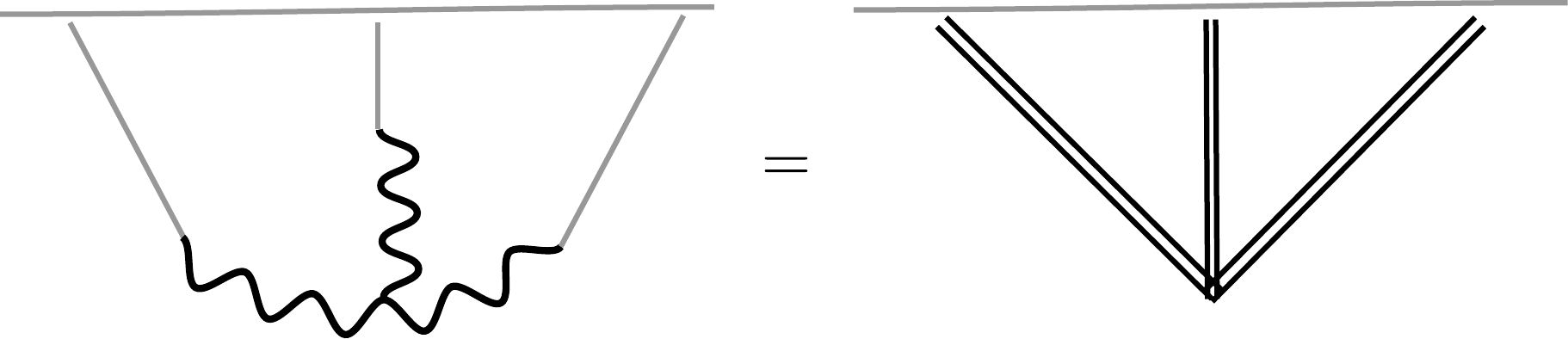}\nonumber
}

To derive this mixed propagator, we use the quadratic mixing from eq.~\eqref{eq:expanded} to write the EOMs satisfied by the Fourier modes $\varphi_{\bf k}$ and $\sigma_{\bf k}$,
\es{coupled_eom}{
&\varphi_{\bf k}''-\frac{2}{\eta}\varphi_{\bf k}'+k^2\varphi_{\bf k}=\rho\left(\frac{1}{\eta}\sigma_{\bf k}'-\frac{3}{\eta^2}\sigma_{\bf k}\right),\\
        &\sigma_{\bf k}''-\frac{2}{\eta}\sigma_{\bf k}'+\left(k^2+\frac{M^2}{\eta^2}\right)\sigma_{\bf k}=-\frac{\rho}{\eta}\varphi_{\bf k}',
}
where $\rho = -2\phidot/\Lambda_1$.
There are two sets of independent solutions to the above equations.
These can be used to canonically quantize the system:
\es{}{
    \begin{pmatrix}\varphi_{\bf k}(\eta)\\\sigma_{\bf k}(\eta)\end{pmatrix}=a^{(1)}_{\bf k}\begin{pmatrix}\varphi_{k}^{(1)}(\eta)\\\sigma_{k}^{(1)}(\eta)\end{pmatrix}+a^{(2)}_{\bf k}\begin{pmatrix}\varphi_{k}^{(2)}(\eta)\\\sigma_{k}^{(2)}(\eta)\end{pmatrix}+{\rm h.c.},
}
where $a^{(1,2)}_{\bf k}$ are annihilation operators.
To determine numerical mode functions $\varphi_{k}^{(1,2)}(\eta), \sigma_{k}^{(1,2)}(\eta)$, we also need the appropriate initial conditions, which we derive by taking the early time $\eta\rightarrow -\infty$ limit of eq.~\eqref{coupled_eom},
\es{coupled_eom_early}{
&\varphi_{\bf k}''-\frac{2}{\eta}\varphi_{\bf k}'+k^2\varphi_{\bf k}-\frac{\rho}{\eta}\sigma_{\bf k}'\approx 0,\\
        &\sigma_{\bf k}''-\frac{2}{\eta}\sigma_{\bf k}'+k^2\sigma_{\bf k}+\frac{\rho}{\eta}\varphi_{\bf k}'\approx 0.
}
These have the solutions,
\es{eq:ic}{
\sigma^{(1)}_k=i \varphi^{(1)}_k = \frac{i}{2k^{3/2}}(-k\eta)^{1+i\rho/2}e^{-i k\eta},\\
\sigma^{(2)}_k=-i\varphi^{(2)}_k = -\frac{i}{2k^{3/2}}(-k\eta)^{1-i\rho/2}e^{-i k\eta}
}
The final mixed propagator is thus given by,
\es{}{
\langle\sigma_{\bf k}(\eta)\varphi_{\bf k}(\eta_f)\rangle'
=\sigma^{(1)}_k(\eta)\varphi^{(1)*}_k(\eta_f)+\sigma^{(2)}_k(\eta)\varphi^{(2)*}_k(\eta_f),
}
with all the four mode functions obtained numerically, with the initial conditions~\eqref{eq:ic}, and $\eta_f\rightarrow 0$ is a time towards the end of inflation.
In terms of these, the final three point function for triple-exchange is given by,
\es{}{
\langle \varphi(\vec{k}_1)\varphi(\vec{k}_2)\varphi(\vec{k}_3)\rangle'=2\kappa~{\rm Re}\int_{-\infty}^0\frac{\dd\eta}{\eta^4}\prod_{j=1}^3\langle\sigma_{{\bf k}_j}(\eta)\varphi_{{\bf k}_j}(0)\rangle'.
}

{\bf Detailed Forms of the Shape Functions for the Chemical Potential Mechanism}---
We evaluate the three point functions using eq.~\eqref{eq:scp_expanded}, by focusing on the two cubic vertices in eq.~\eqref{eq:scp_cubic} separately.
Since these contributions involve single-exchange processes, we can obtain the full shapes analytically using cosmological bootstrap techniques.
In particular, we follow the convention of~\cite{Qin:2023ejc} where closed form expressions for several single-exchange processes were given in terms of a seed integral,
\begin{widetext}
\es{}{
\mathcal{I}_{\pm\pm}^{p_1,p_2}(u,1) = \frac{\pm i e^{\mp i(p_1+p_2)\pi/2}}{2^{7/2+p_2}\pi^{1/2}} {\Gamma(\frac{5}{2}+p_2-i\nu)\Gamma(\frac{5}{2}+p_2+i\nu)\over \Gamma(3+p_2)}\left(e^{\pi\nu}\mathcal{Y}_{\pm}^{p_1}(u) + e^{-\pi\nu}\mathcal{Y}_{\mp}^{p_1}(u)\right)
\\
+ \frac{e^{\mp i(p_1+p_2)\pi/2}\Gamma(5+p_1+p_2)\,u^{5+p_1+p_2}}{2^{5+p_1+p_2}\left[\left(\frac{5}{2}+p_2\right)^2+\nu^2\right]} \,{}_3F_2\!\left[\begin{array}{c} 1,\ 3+p_2,\ 5+p_1+p_2 \\ \frac{7}{2}+p_2-i\nu,\ \frac{7}{2}+p_2+i\nu \end{array}\bigg| u\right],
}
\es{}{
\mathcal{I}_{\pm\mp}^{p_1,p_2}(u,1) = \frac{ e^{\mp i(p_1-p_2)\pi/2}}{2^{7/2+p_2}\pi^{1/2}} {\Gamma(\frac{5}{2}+p_2-i\nu)\Gamma(\frac{5}{2}+p_2+i\nu)\over \Gamma(3+p_2)}\left(\mathcal{Y}_{+}^{p_1}(u) + \mathcal{Y}_{-}^{p_1}(u)\right),
}
\es{}{
\mathcal{Y}_{\pm}^{p}(u) = 2^{\mp i\nu} \left(\frac{u}{2}\right)^{5/2+p\pm i \nu} \Gamma(5/2+p\pm i\nu)\Gamma( \mp i\nu) {}_2F_1\!\left[\begin{array}{c} \frac{5}{2}+p\pm i\nu,\ \frac{1}{2}\pm i\nu \\ 1\pm 2i\nu \end{array}\bigg| u\right],
}
\end{widetext}
where $u=2k_3/(k_1 + k_2 + k_3)$.

For the cubic coupling given in the first and second of eq.~\eqref{eq:scp_cubic}, we obtain the the three point functions respectively as (with $p_2 = -2+i\omega$),
\begin{widetext}
\es{eq:CPtime}{
&\langle \varphi(\vec{k}_1)\varphi(\vec{k}_2)\varphi(\vec{k}_3)\rangle'\rvert_{\rm temp} \\
&=  {i \over 8 k_1 k_2 k_3}\left(\frac{2c\omega}{\Lambda}\right)^2 \frac{|\chi_0|^2}{\Lambda} \times \sum_{a,b=\pm} \left[\left(\frac{1}{k_2^2k_3}+\frac{1}{k_1^2 k_3}\right)\mathcal{I}_{ab}^{-2-i \omega, p_2}(u,1)+i a \left(\frac{1}{k_2 k_3^2}+\frac{1}{k_1 k_3^2}\right)\mathcal{I}_{ab}^{-1-i \omega, p_2}(u,1)-(\omega\rightarrow -\omega)\right]\\
&+{\text{cyclic perms.,}}}
\es{cp_shape}{
&\langle \varphi(\vec{k}_1)\varphi(\vec{k}_2)\varphi(\vec{k}_3)\rangle'\rvert_{\rm cov} \\&= -{1\over 8 k_1k_2k_3^4}\left(\frac{2c}{\Lambda}\right)^2\frac{\omega|\chi_0|^2}{\Lambda}\sum_{a,b=\pm}\left[\left({\cal I}_{ab}^{-i\omega,p_2}+{k_3^2\left(\vec{k}_1\cdot\vec{k}_2\right)\over k_1^2k_2^2}\left({\cal I}_{ab}^{-2-i\omega,p_2}+i a{k_{12}\over k_3}{\cal I}_{ab}^{-1-i\omega,p_2}-{k_1k_2\over k_3^2}{\cal I}_{ab}^{-i\omega,p_2}\right)\right)+(\omega\rightarrow-\omega)\right]\\
&+{\text{cyclic perms.}}.
}
\end{widetext}

\bibliography{references_letter}

\end{document}